\documentclass[reqno,12pt,a4paper]{amsart}

\usepackage{graphicx}
\usepackage{psfrag}

\usepackage{mathrsfs}
\usepackage{amssymb}
\usepackage{amsfonts}
\usepackage{latexsym}
\usepackage{amsmath}
\usepackage{amsthm}

\newcommand{\er}[1]{\textrm{(\ref{#1})}}
\def\lb{\label}

\theoremstyle{plain}

\newtheorem{theorem}{\bf Theorem}[section]
\newtheorem{lemma}[theorem]{\bf Lemma}

\theoremstyle{remark}
\newtheorem{remark}[theorem]{\bf Remark}

\renewcommand{\a}{\alpha}           \newcommand{\cA}{\mathcal{A}}

            \newcommand{\cB}{\mathcal{B}}

\renewcommand{\d}{\delta}

\newcommand{\ve}{\varepsilon}     \newcommand{\cG}{\mathcal{G}}

\renewcommand{\l}{\lambda}          

\renewcommand{\L}{\Lambda}

\newcommand{\s}{\sigma}

\renewcommand{\t}{\tau}             

\newcommand{\cU}{\mathcal{U}}

\newcommand{\F}{\Phi}

             \newcommand{\cZ}{\mathcal{Z}}
      
\renewcommand{\o}{\omega}

\def\Z{\mathbb{Z}}
\def\R{\mathbb{R}}
\def\C{\mathbb{C}}
\def\T{\mathbb{T}}
\def\N{\mathbb{N}}

\def\qqq{\qquad}
\def\qq{\quad}

\let\ge\geqslant
\let\le\leqslant

\newcommand{\ca}{\begin{cases}}
\newcommand{\ac}{\end{cases}}
\newcommand{\ma}{\begin{pmatrix}}
\newcommand{\am}{\end{pmatrix}}

\def\lt{\biggl}
\def\rt{\biggr}

\renewcommand{\[}{\begin{equation}}
\renewcommand{\]}{\end{equation}}

\def\wt{\widetilde}
\def\pa{\partial}
\def\sm{\setminus}

\def\no{\noindent}
\def\ol{\overline}
\def\iy{\infty}

\def\/{\over}
\def\ts{\times}

\def\Re{\mathop{\rm Re}\nolimits}

\def\diag{\mathop{\rm diag}\nolimits}

\def\BBox{\hspace{1mm}\vrule height6pt width5.5pt depth0pt \hspace{6pt}}
\def\wh{\widehat}
\def\as{\text{as}}
\def\where{\text{where}}
\def\1{1\!\!1}

\begin{document}

\title [Sharp eigenvalue  asymptotics for
 fourth order operators on the circle]
 {Sharp eigenvalue asymptotics for
 fourth order operators on the circle}

\date{\today}
\author[Andrey Badanin]{Andrey Badanin}
\address{Northern (Arctic) Federal University,
Northern Dvina emb, 17, Arkhangelsk, 163002, Russia, e-mail:
an.badanin@gmail.com}
\author[Evgeny Korotyaev]{Evgeny Korotyaev}
\address{Mathematical Physics Department, Faculty of Physics, Ulianovskaya 2,
St. Petersburg State University, St. Petersburg, 198904,
 Russia,
 \ korotyaev@gmail.com,}

\maketitle

\begin{abstract}
We determine high energy asymptotics of eigenvalues of
fourth order operator on the circle.
\end{abstract}

\section {Introduction and main results}
\setcounter{equation}{0}

We consider the operator $H$ acting on the circle $2(\R/\Z)$ and
given by
\[
\lb{fo} H=\pa^4+2\pa p\pa +q.
\]
i.e, we consider operator $H$ on the interval $[0,2]$ with the
2-periodic boundary conditions. We assume that $p,q$ are the  real
1-periodic functions, which satisfy the conditions:
\[
\lb{pq}
p, q\in L^1(\T),\qq\T=\R/\Z,\qq\int_0^1q(t)dt=0.
\]
It is well known that the operator $H$ is self-adjoint and its
spectrum consists of eigenvalues of multiplicity $\le 4$.
Really, eigenvalues of multiplicity 4 exist,
see Remarks \ref{rem1}.

Let
$\l_0^+,\l_{n}^\pm,n\in\N$, be the eigenvalues of the operator $H$
labeled by
$$
\l_0^+\le\l_1^-\le\l_1^+\le\l_2^-\le...,
$$
counted with multiplicities. These
eigenvalues satisfy the asymptotics
\[
\lb{4g.T2-2}
\l_n^{\pm}=(\pi n)^4
+2(\pi n)^2\big(- p_0\pm |\wh p_n|+O(n^{-{1\/2}})\big)
\]
as $n\to+\iy$, see \cite{BK1}, where
$$
f_0=\int_0^1f(t)dt,\qqq
\wh f_n=\int_0^1 f(t)e^{-i2\pi nt}dt\qqq\forall\qq n\in\Z.
$$
There are no coefficients $\wh q_n$
in the leading term of asymptotics \er{4g.T2-2}, since $p, q$
satisfy \er{pq}. Our goal is to determine  asymptotics of eigenvalues, when
there are the coefficients $\wh q_n$ in the leading term. In this case we need
an additional condition $p''\in L^2(\T)$.

\subsection{Second order operators}
Firstly we recall the well-known  asymptotics of eigenvalues for
the second order operator $-\pa^2+q$ on the circle.
The spectral properties of  Schr\"odinger operators $-\pa^2+q$ on $\T$
with the periodic $q$ are well understood, see, e.g., the books
of Levitan--Sargsyan \cite{LeS}, Magnus--Winkler \cite{MW}, Marchenko \cite{M},
Titchmarsh \cite{T}, and references therein.
Consider the second order operator $-\pa^2+q$ on the interval
$[0,2]$ with the
2-periodic boundary conditions.
If $q\in C^\iy(\T)$, then the
eigenvalues $\a_0^+,\a_n^\pm,n\in\N$, of this operator have
multiplicity 1 or 2 and satisfy the inequalities
$\a_0^+<\a_1^-\le\a_1^+<\a_2^-\le...$ and the asymptotics
\[
\lb{an}
\a_n^\pm=(\pi n)^2+{c_0\/(\pi n)^2}+{c_1\/(\pi n)^4}+....
\]
Here the numbers $c_j$ are expressed in terms of $q$ and the derivatives
$q^{(s)}, s\le j$, see \cite[Th~1.5.2]{M}.
The full asymptotic expansion for the case
$q,q^{(m)}\in L^2(\T)$, where $m\in\N$, was determined by
Marchenko--Ostrovski \cite{MO}.
The case, when the potential $q$ is a distribution, $q=u', u\in L^2(\T)$,
was considered by Korotyaev \cite{K2}.

\subsection{Fourth  order operators}
The standard applications of the fourth order differential equations
are bending vibrations of thin beams and plates described
by the Euler-Bernoulli equation, see \cite{Gl}.
Moreover, these equations arise in many physical models:
hydrodynamic stability (Orr--Sommerfeld  equation, \cite{Li}),
kinetic of liquid phase (Cahn--Hilliard equation, \cite{CH}),
elastic buckling \cite{HW}, thin films \cite{BGW}, see also
the book \cite{PeT} and references therein.

Many papers are devoted to the study of the
spectral problems for the fourth order operators,
see papers
Caudill--Perry--Schueller \cite{CPS}, McLaughlin \cite{McL},
Hoppe--Laptev--\"Ostensson \cite{HLO},
Mikhailets--Molyboga \cite{MMo},
Papanicolaou \cite{P1}, \cite{P2},
Badanin--Korotyaev \cite{BK1} and references therein.

Caudill, Perry and Schueller \cite{CPS} described the
so-called iso-spectral potentials for the operator $\pa^4+2\pa p\pa +q$ with the
boundary condition $f(0)=f''(0)=f(1)=f''(1)=0$.

McLaughlin \cite{McL} studied the inverse spectral
problems by the spectrum and the normalization constants
for the operator $\pa^4+\pa p\pa +q$ on the
interval $[0,1]$ with the boundary conditions
$f(0)=f'(0)=f(1)=f'(1)=0$.

Hoppe, Laptev and \"Ostensson \cite{HLO} considered the inverse
spectral problem for the operator $\pa^4+\pa p\pa +q$ on the line
in the case of rapidly decaying $p,q$ at infinity.

Mikhailets and Molyboga \cite{MMo} considered
the operator $\pa^4+q$ on the circle, where
the function $q$ is a distribution. They determined asymptotics of eigenvalues  for this operator.

In the series of papers Papanicolaou \cite{P1}-\cite{P3} and jointly with
Kravvaritis \cite{PK1}-\cite{PK2} the Euler-Bernoulli operator
${1\/b}(af'')''$ on the line with the periodic positive $a,b$ is considered.
The unitary Liouville's type transformation reduces this operator
to the operator $\pa^4+\pa p\pa +q$ with some special periodic $p,q$.
The spectral properties of the Euler-Bernoulli operator are simpler, than for
the general operator $\pa^4+\pa p\pa +q$.
The spectrum lies on $\R_+$ and its structure is similar
to the structure of the spectrum of the Hill operator:
the spectrum has multiplicity 2, consists of non-degenerated
intervals separated by gaps, the endpoints of gaps
are eigenvalues of the 2-periodic problem.
The necessary and sufficient conditions for the Euler-Bernoulli
operator to be a perfect square of the operator $-af''$
are found in \cite{P3}.

Badanin and Korotyaev \cite{BK1} considered the operator
$\pa^4+\pa p\pa +q$ with periodic $p,q$ on the line.
The spectrum of this operator consists of intervals,
separated by gaps, has multiplicity 2 or 4.
Authors determine asymptotics of the gaps.
This asymptotics shows that
the spectrum has multiplicity 2 at high energy and for
the generic coefficients there exists an infinite number of gaps.

Spectral asymptotics for the higher order operators are much less
investigated. Numerous results about the regular and singular
boundary value problems for these operators are expounded in the
book of Naimark \cite{Na}. High energy asymptotics of the spectral gaps
for even order operator with periodic coefficients
is determined by Badanin and Korotyaev \cite{BK2}. The scattering theory for
higher order operators is a subject of the book
of Beals, Deift, Tomei \cite{BDT}.

\subsection{Main results}
 In order to determine asymptotics including the term $\wh
q_n$ we need an additional condition $p''\in L^1(\T)$, and finally
we have also consider the case ${p'''}, q'\in L^1(\T)$.

\begin{theorem}
\lb{4g.thaspi}
i) If $p,p'',q\in L^1(\T)$,
then the eigenvalues of $H$
satisfy
\[
\lb{4g.T2-2i} \l_n^\pm=(\pi n)^4-(\pi n)^2
p_0-{\|p\|^2-p_0^2\/2}\pm|\wh V_n|+{o(1)\/n^{1\/2}} \qqq\as\qq
n\to+\iy,
\]
where
$$
V=q-{p''\/2},\qqq \|p\|^2=\int_0^1|p(t)|^2dt.
$$

ii) Let, in addition, $p''', q'\in L^1(\T)$. Then the eigenvalues of
$H$ satisfy
\[
\lb{4g.T2-2ii} \l_n^\pm=(\pi n)^4-(\pi n)^2 p_0-{\|p\|^2-
p_0^2\/2}\pm|\wh V_n| +{o(1)\/n^{3\/2}}\qqq\as\qq n\to+\iy.
\]
\end{theorem}

\begin{remark}
\lb{rem1}
i) Consider the operator $(-\pa^2-10\pi^2)^2$. This operator
has an 1-periodic eigenvalue $36\pi^4$ of multiplicity 4.
The corresponding eigenfunctions have the form
$e^{\pm i2\pi t},e^{\pm i4\pi t}$.

ii) Erovenko \cite{E} considered
the operator \er{fo} and determined the
following asymptotics for its eigenvalues
\[
\lb{asE}
\l_n^+-\l_n^-=2\Big(|\wh q_n|^2+{|\wh
p_n''|^2\/4}\Big)^{1\/2}\big(1+o(1)\big)\qqq\as\qq n\to+\iy.
\]
Unfortunately, this asymptotics
is not correct, see more in Section 4.

iii) We use asymptotics \er{4g.T2-2ii}
in order to obtain the trace formula for
the operator $H$ \cite{BK3}.

iv) There is an open problem: to determine the full asymptotic
expansion in the case $p,q\in C^\iy(\T)$.
\end{remark}

\subsection{}
Describe briefly our proofs. We rewrite the initial differential
equation \er{4g.1b} into the matrix form \er{4g.Me}. The $4\ts 4$
matrix solution $M(t,\l)$ of equation \er{4g.Me} satisfying the
condition $M(0,\l)=\1_4$ is the {\it fundamental matrix}. The matrix
$M(1,\l)$ is the {\it monodromy matrix}. Our main tool is an
analysis of this matrix. Such analysis for the Schr\"odinger
operator with the periodic matrix potential was made by Chelkak,
Korotyaev \cite{CK}. But in our case we meet additional
difficulties. For the second order operators (even with the matrix
coefficients) all  entries of the monodromy matrix are bounded for
$\l\to+\iy$ (in the unperturbed case they have the form
$\cos\sqrt\l,\sin\sqrt\l$). For
the fourth order differential operator some entries of the monodromy
matrix are bounded as $\l\to+\iy$ and all other entries are
unbounded
(in the unperturbed case they have the form
$\cos\l^{1\/4},\sin\l^{1\/4},\cosh\l^{1\/4},\sinh\l^{1\/4}$).

Asymptotics of eigenvalues for the higher order operators in the
case of smooth coefficients was determined by Birkhoff (see
\cite[Ch.~II.4]{Na}). Here we consistently apply the matrix form of
Birkhof's method (the similar method was used in \cite{BK2}). The
calculations in the matrix form are simpler than in the original
scalar form. Note that the matrix approach was used before by
Chelkak, Korotyaev \cite{CK} for the second order operator with the
periodic matrix potential. However, in the case \cite{CK} the situation is
simpler, since the monodromy matrix is bounded at $\l\to+\iy$.
Roughly speaking the determining of asymptotics for
higher order operators has the difficulties of
analysis of the systems plus additional difficulties from
the increasing solutions.

In order to overcame these problems we need some modifications
of the matrix differential equation, see Lemma \ref{lm21}.
Following to Fedoruk \cite[Ch.~V.1.3]{Fe}, Korotyaev \cite{K1},
we modify equation \er{4g.Me} to the quasi-diagonal form,
where the matrix coefficient of the equation is the sum of the
diagonal matrix and the small perturbation as $|\l|\to\iy$,
see \er{4g.eqwM}. After that, following to Birkhoff (see
\cite[Ch.~II.4]{Na}), we rewrite the matrix differential
equation into the form of an integral equation, and show that this equation
is uniquely solvable for all large $|\l|$. Finally we
obtain the special representation of the monodromy
matrix in the form of the multiplication of the simple diagonal
matrix and the matrix $F$ bounded for all large $|\l|$, see
\er{4g.reprM}. Iterations of the integral equation
allow us to determine the asymptotics of the matrix $F$,
and then asymptotics \er{4g.T2-2i}, \er{4g.T2-2ii}.

The plan of the paper is as follows. In Section 2 we transform the
differential equation for the fundamental matrix to the quasi-diagonal form.
In Section 3 we reduce the matrix differential equation to the
equivalent integral equation. Iterations of this integral
equation give us the asymptotics of the fundamental matrix.
In Section 4 we analyze the characteristic determinant of the
monodromy matrix and determine the asymptotics of eigenvalues
of $H$.

\section {Fundamental matrix}
\setcounter{equation}{0}

Consider the equation
\[
\lb{4g.1b}
f^{(4)}+2(pf')'+qf=\l f
\]
on $\R$, where $\l\in\C$. Rewrite equation \er{4g.1b} in the matrix
form
\[
\lb{4g.fe}
{\bf f}'-\L {\bf f}=-(2pJ+qJ_1){\bf f},
\]
where the vector ${\bf f}(t)$ and the matrices
$\L(\l),J,J_1$ are given by
\[
\lb{4g.LQ}
{\bf f}=\ma f\\ f'\\f''\\f'''+2pf\am,
\qq
\L=\ma
0&1&0&0\\
0&0&1&0\\
0&0&0&1\\
\l&0&0&0\\
\am,
\qq
J=\ma
0&0&0&0\\
0&0&0&0\\
0&1&0&0\\
0&0&0&0\\
\am,\qq
J_1=\ma
0&0&0&0\\
0&0&0&0\\
0&0&0&0\\
1&0&0&0\\
\am.
\]
Define the $4\ts 4$~-~matrix valued solution $M(t,\l)$
of equation \er{4g.fe}:
\[
\lb{4g.Me}
M'-\L M=-(2pJ+qJ_1)M,\qqq M(0,\l)=\1_4,
\]
where $\1_4$ is the identity $4\ts 4$~-~matrix. $M(t,\l)$ is called
the {\it fundamental matrix} of equation \er{4g.1b}. Each
function $M(t,\cdot), t\in\R$, is entire and real on $\R$.
The matrix valued function $M(1,\l)$ is called the {\it
monodromy matrix}.
The spectrum $\s(H)$ of the operator $H$
satisfies the identity
$$
\s(H)=\{\l_0^+,\l_n^\pm,n\ge 1\}=\{\l\in\R:\det(M(1,\l)\pm \1_4)=0\}.
$$

Introduce the unitary  $4\ts 4$~-~matrices
$\o$ and $U$ given by
\[
\lb{4g.Om}
\o=\diag(\o_1,\o_2,\o_3,\o_4)=\diag(i,1,-1,-i),
\]
$$
U={1\/2}(\o_k^{j-1})_{j,k=1}^4={1\/2}\ma 1&1&1&1\\
i&1&-1&-i\\
-1&1&1&-1\\
-i&1&-1&i\am.
$$
Define the $4\ts 4-$matrices
\[
\lb{4g.idA} A=U^{-1}J
U={1\/4}\ma-i&-1&1&i\\i&1&-1&-i\\i&1&-1&-i\\-i&-1&1&i\am, \qq
A_1=-U^{-1}J_1
U={1\/4}\ma-i&-i&-i&-i\\-1&-1&-1&-1\\1&1&1&1\\i&i&i&i\am,
\]
\[
\lb{4g.idB}
B={1\/8}\ma0&1+i&1-i&1\\-1+i&0&-1&-1-i\\
-1-i&-1&0&-1+i\\ 1&1-i&1+i&0\am,\qq
B_1={1\/16}\ma 0&-2&-2&-1\\2i&0&i&2i\\
-2i&-i&0&-i\\1&2&2&0\am,
\]
\[
\lb{4g.O-AB}
B_2={1\/32}\ma i&i&i&0\\ 1&1&0&1
\\-1&0&-1&-1\\0&-i&-i&-i\am.
\]

Introduce the variable $z=\l^{1\/4},\l\in\C,$ by the condition
$$
\arg z\in\Big(-{\pi\/4},{\pi\/4}\Big]\qqq\as\qq\arg\l\in[0,2\pi).
$$
If $\arg\l\in[0,\pi)$, then
$$
z\in S=\Big\{z\in\C:\arg z\in\big[0,{\pi\/4}\big)\Big\},
$$
and we have the following estimates:
\[
\lb{4g.esom}
\Re(i\o_1z)\le\Re(i\o_2z)\le\Re(i\o_3z)\le\Re(i\o_4z)
\qqq\forall\qq z\in S.
\]

Introduce the $4\ts 4$~-~matrix valued functions
\[
\lb{4g.Z}
\cZ=\diag\big(1,iz,(iz)^2,(iz)^3\big),
\]
\[
\lb{4g.cS}
 W(t,z)=\1_4-{2p(t)\/z^2}B-{2p'(t)\/z^3}B_1,
\]
$(t,z)\in\R\ts S.$
In the following Lemma we modify equation
\er{4g.Me} into the quasi-diagonal form \er{4g.eqwM}.

\begin{lemma}
\lb{lm21}
Let
$$
z\in S_r=\{z\in S:|z|>r\}
$$
for some $r>0$ large enough.
Then the matrix valued function $\F (t,z),t\in\R$, given by the identity
\[
\lb{4g.MwM}
\F (t,z)=\big(\cZ(z) U W(t,z)\big)^{-1}M(t,z^4)\cZ(z) U W(0,z),
\]
satisfies the equation on the real line
\[
\lb{4g.eqwM}
\F '-iz\xi\F =Q\F,\qqq\F (0,z)=\1_4,
\]
where
\[
\lb{4g.wtO}
\begin{aligned}
\xi(t,z)=\o+{p(t)\/2z^2}\o^*=\diag(\xi_j(t,z))_{j=1}^4,
\\
\xi_j(t,z)=\o_j+{\ol\o_jp(t)\/2z^2}\qqq \forall\qq j\in\N_4=\{1,2,3,4\},
\end{aligned}
\]
\[
\lb{4g.wtQ}
Q(t,z)={Q_0(t)\/z^3}+{4p(t)p'(t)Q_1\/z^4}+{Q_2(t,z)\/z^5},
\]
\[
\lb{4g.wtQ0}
Q_0(t)=iq(t)A_1+2p''(t)B_1+i4p^2(t)B_2,
\]
\[
\lb{Q1}
Q_1=-iAB_1+B\big(B-i\o B_1\big)+iB_1\big(A-\o B),
\]
the matrix valued function $Q_2(t,z)$, given by \er{Q2},
is uniformly bounded on $\R\ts S_r$, and each function
$Q_2(t,\cdot),t\in\R$, is analytic in $S_r$.

\end{lemma}

\no {\bf Proof.}
Multiplying equation \er{4g.Me} by the matrix
$\cZ U$ from the right
and by the matrix $(\cZ U)^{-1}$ from the left
and using the identities
\begin{gather*}
(\cZ U)^{-1}\L\cZ U=\o,
\\
-(\cZ U)^{-1}\big(2pJ+qJ_1\big)\cZ U
={i\/z} U^{-1}\big(2pJ-{q\/z^2}J_1\big) U
={i2p\/z}A+{iq\/z^3}A_1,
\end{gather*}
we obtain the equation
\[
\lb{4g.me2}
\wt\Phi'-iz\o\wt\Phi=\big({i2p\/z}A+{iq\/z^3}A_1\big)\wt\Phi,
\]
with respect to the matrix valued function
$
\wt\Phi=(\cZ U)^{-1}M\cZ U.
$

Identity \er{4g.MwM} gives
$$
\wt\Phi(t,z)= W(t,z)\F(t,z)  W^{-1}(0,z).
$$
Substituting this identity into \er{4g.me2}, we obtain
$$
 W\F '=
\big(iz\o W- W'+{i2p\/z}A W+{iq\/z^3}A_1 W\big)\F ,
\qqq W= W(t,z).
$$
Substituting \er{4g.cS} into the last identity we obtain
\[
\lb{4g.cScM}
 W\F '=
\Big(iz\o +{i2p\/z}\big(A-\o B)+{2p'\/z^2}\big(B-i\o B_1\big)
+{Q_0\/z^3}+{\wt A\/z^4}\Big)\F ,
\]
where
\[
\lb{4g.cCpr}
Q_0=i4p^2[B,A]+i4p^2B[B,\o]+iqA_1+2p''B_1,\qqq
[\cA,\cB]=\cA\cB-\cB\cA,
\]
\[
\lb{wtA}
\wt A=-i4pp'AB_1-i{2pq\/z}A_1B-i{2p'q\/z^2}A_1B_1,
\]
Introduce the matrix valued function $\wt B(t,z),(t,z)\in\R\ts S_r$, by the identity
\[
\lb{W^-1}
W^{-1}(t,z)=\1_4+{2p(t)\/z^2}B+{2p'(t)\/z^3}B_1+{\wt B(t,z)\/z^4}.
\]
Then $\wt B(t,z)$ is uniformly bounded on $\R\ts S_r$ and
$B(t,\cdot)$ is analytic in $S_r$ for all $t\in\R$.

Substituting \er{W^-1} into \er{4g.cScM} we obtain
\[
\lb{ur1}
\F '=\Big(iz\o+{i2p\/z}([B,\o]+A)+{2p'\/z^2}(B+i[B_1,\o])
+{Q_0\/z^3}+{4pp'Q_1\/z^4}+{Q_2\/z^5}\Big)\F ,
\]
where the constant matrix $Q_1$ is given by \er{Q1},
\[
\lb{Q2}
\begin{aligned}
Q_2=-i2qA_1\Big(pB+{p'B_1\/z}\Big)+2pB\Big(Q_0+{\wt A\/z}\Big)
+2p'B_1\Big(2p'\big(B-i\o B_1\big)
+{Q_0\/z}+{\wt A\/z^2}\Big)
\\
+\wt B\Big(i2p\big(A-\o B)+{2p'\/z}\big(B-i\o B_1\big)
+{Q_0\/z^2}+{\wt A\/z^3}\Big).
\end{aligned}
\]
This identity shows that $Q_2(t,z)$ is uniformly bounded on $\R\ts S_r$ and
each function $Q_2(t,\cdot),t\in\R$, is analytic in $S_r$.

The matrices $B,B_1,B_2$ satisfy the identities
$$
[B,\o]+A={\o^*\/4},\qqq B+i[B_1,\o]=0,\qqq B_2={B\o^*\/4}-AB.
$$
Substituting these identities into \er{ur1} and \er{4g.cCpr} we obtain \er{4g.eqwM}
and \er{4g.wtQ0}.
$\BBox$

\section{Asymptotics of the monodromy matrix}
\setcounter{equation}{0}

Let $r>0$ be large enough and let $z\in S_r$.
In equation \er{4g.eqwM} we introduce
the new unknown $4\ts 4$~-~matrix valued function
$G(t,z),t\in[0,1]$, given by
the identity
\[
\lb{4g.rmmi}
\F (t,z)=G(t,z) e^{i z\int_0^t\xi(s,z)ds}G^{-1}(0,z).
\]
Substituting \er{4g.rmmi} into \er{4g.eqwM}
we obtain the matrix differential equation
on the interval $[0,1]$
\[
\lb{4g.ecGi}
G'+iz(G\xi-\xi G)=QG.
\]
If the matrix valued function $G$ satisfies equation \er{4g.ecGi},
then the matrix valued function $\F$, given by \er{4g.rmmi},
satisfies \er{4g.eqwM} on the interval $[0,1]$.

We rewrite equation \er{4g.ecGi} in the form of the matrix
integral equation on the interval $[0,1]$
\[
\lb{4g.me5i}
G=\1_{4}+KG,
\]
where
\[
\lb{4g.dcLi}
(KG)_{\ell j}(t,z)=\int_0^1K_{\ell j}(t,s,z)(QG)_{\ell j}(s,z)ds
\qqq\forall\qq\ell ,j\in\N_{4},
\]
\[
\lb{Klj}
K_{\ell j}(t,s,z)=
\ca \ \  e^{iz\int_s^t(\xi_\ell(u,z)-\xi_j(u,z))du}\chi(t-s),
\ \ \  \ell <j\\
-e^{iz\int_s^t(\xi_\ell(u,z)-\xi_j(u,z))du}\chi(s-t), \ \ \
\ell \ge j\ac,\qq \chi(s)=\ca 1,\ s\ge 0\\ 0,\ s<0\ac.
\]

In the following Lemma we will show that equation \er{4g.me5i} for
$z\in S_r$
has the unique solution in the class of
$4\ts 4$~-~matrix valued functions from
$L^\iy(0,1)$ with the norm given by
$$
\|\cA\|_\iy^2=\sup_{t\in[0,1]}|\cA(t)|^2,\qqq
|\cA|^2=\max_{h\in\C^4:\sum |h_j|^2=1}\sum_{j=1}^4|(\cA h)_j|^2.
$$

\begin{lemma}
\lb{4g.lmwtG}
Let $z\in S_r$ for some $r>0$ large enough.
Then equation \er{4g.me5i}
has the unique solution $G(\cdot,z)\in L^\iy(0,1)$.
Moreover,
\[
\lb{4g.me6i}
G(t,z)=\sum_{n=0}^\iy G_{n}(t,z)\qqq\forall\qq(t,z)\in[0,1]\ts S_r,
\]
where
\[
\lb{4g.itGi}
G_{0}=\1_{4},\qqq
G_{n}=KG_{n-1}=K^n\1_4\qqq\forall\qq n\in\N.
\]
The series \er{4g.me6i}
converges absolutely and uniformly on any compact set in
$[0,1]\ts S_r$.
Each matrix valued function $G(t,\cdot),t\in[0,1]$,
is analytic in $S_r$ and it satisfies the asymptotics
\[
\lb{4g.aswtG}
G(t,z)=\1_{4}+O(z^{-3}),\qqq
G(t,z)=\1_{4}+G_1(t,z)+O(z^{-6})
\]
as $|z|\to\iy,z\in S$, uniformly on $t\in[0,1]$.
\end{lemma}

\no {\bf Proof.}
Identity \er{4g.wtO} gives
$$
\xi_\ell(t,z) -\xi_j(t,z)=\o_\ell-\o_j+{p(t)\/2z^2}(\ol\o_\ell-\ol\o_j)
\qq\forall\qq \ell,j\in\N_4.
$$
Then estimates \er{4g.esom} yield
$$
\Re iz\big(\xi_\ell(t,z)-\xi_j(t,z)\big)
=\Re iz(\o_\ell-\o_j)+{p(t)\/2}\Re {i(\ol\o_\ell-\ol\o_j)\/z}
\le{|p(t)|\/|z|}
$$
for all $t\in[0,1],\ell,j\in\N_4:\ell\le j$.
Substituting this estimate into \er{Klj} we obtain
\[
\lb{4g.estkerK} |K_{\ell j}(t,s,z)| \le e^{\|p\|\/|z|}\le 2
\]
for all $t,s\in[0,1],\ell,j\in\N_4,z\in S_r$.

Let the $4\ts 4$~-~matrix valued function $\cA$ belongs to the class
$L^\iy(0,1)$.
Substituting estimates \er{4g.estkerK} into
\er{4g.dcLi},
we obtain
\begin{multline*}
\|(K\cA)(\cdot,z)\|_\iy=
\sup_{t\in[0,1]}|(K\cA)(t,z)|
\le4\sup_{t\in[0,1]}\max_{(i,j)\in\N_{4}^2}|(K\cA)_{ij}(t,z)|
\\
\le8\max_{(i,j)\in\N_{4}^2}\int_0^1|(Q\cA)_{ij}(s,z)|ds
\le8\int_0^1|Q(s,z)||\cA(s)|ds
\le8\|\cA\|_\iy\int_0^1|Q(s,z)|ds.
\end{multline*}
Identity \er{4g.wtQ} gives
$$
\int_0^1|Q(s,z)|ds\le{C\/|z|^3} \qqq\forall\qq (t,z)\in[0,1]\ts S_r
$$
for some $C>0$.
Therefore,
\[
\lb{4g.KKi}
\|(K\cA)(\cdot,z)\|_\iy\le {8C\/|z|^3}\|\cA\|_\iy\qqq\forall\qq z\in S_r.
\]

Iterations in \er{4g.me5i} give \er{4g.me6i}.
Estimate \er{4g.KKi} yields
\[
\lb{4g.me7i} \|G_{n}(\cdot,z)\|_\iy
=\|(K^n\1_4)(\cdot,z)\|_\iy\le\Big({8C\/|z|^3}\Big)^n\qqq \forall\qq
(n,z)\in\N\ts S_r.
\]
These estimates show that the formal series \er{4g.me6i}
converges absolutely and uniformly on any compact in $S_r$.
Therefore, it gives the unique solution of equation \er{4g.me5i}.
Each term of this series is analytic with respect to $z$ in $S_r$.
Therefore, the function $G$ is analytic also.
Substituting estimates \er{4g.me7i} into series \er{4g.me6i}
we obtain asymptotics \er{4g.aswtG}.
$\BBox$

Identity \er{4g.rmmi} yields
\[
\lb{4g.wMGF}
\F (1,z)=G(0,z)F(z) e^{i zv(z)}G^{-1}(0,z)\qqq\forall\qq z\in S_r,
\]
where the $4\ts 4$~-~matrix valued functions $F(z),v(z)$ are given by
\[
\lb{4g.wtF}
F(z)=G^{-1}(0,z)G(1,z),\qqq
v(z)=\int_0^1\xi(t,z)dt.
\]
Identities \er{4g.Om}, \er{4g.wtO} yield
\[
\lb{defv}
v(z)=\int_0^1\Big(\o+{p(t)\/2z^2}\o^*\Big)dt=\diag( v_j(z))_{j=1}^4,\qqq
v_j(z)=\o_j+{p_0\/2z^2}\ol\o_j.
\]
Lemma \ref{4g.lmwtG} shows that
$F(z)$ is analytic and uniformly bounded on $S_r$.

Identities \er{4g.MwM} and \er{4g.wMGF} give the following
basic representation for the monodromy matrix
\[
\lb{4g.reprM}
M(1,z^4)=\cU(z)F(z) e^{i zv(z)}
\cU^{-1}(z)\qqq\forall\qq z\in S_r,
\]
where
$$
\cU(z)=\cZ(z) U W(0,z)G(0,z).
$$

In the following Lemma we determine asymptotics of the matrix
valued function $F$.

\begin{lemma}
i) The function $F(z)$ satisfies the asymptotics
\[
\lb{4g.aswcF}
F_{jk}(z)=O(z^{-3})\qqq\forall\qq j,k\in\N_4: j\ne k,
\]
\[
\lb{4g.wtf2i}
F_{j j}(z)=1+{i\o_j\|p\|^2\/8z^3}
+{O(1)\/z^5}\qqq\forall\qq j\in\N_4
\]
as $|z|\to\iy,z\in S$,
\[
\lb{4g.asF3223}
F_{2 3}(z)=-{i\wh V_n\/4(\pi n)^3}
+{o(1)\/n^4},\qq
F_{ 32}(z)
={i\ol{\wh V_n}\/4(\pi n)^3}+{o(1)\/n^4}
\]
as $z=(\l_n^\pm)^{1\/4},n\to+\iy$.

ii) Let $p,p''',q,q'\in L^1(\T)$. Then
\[
\lb{4g.asF3223i}
F_{2 3}(z)=-{i\wh V_n\/4(\pi n)^3}
+{O(1)\/n^5},\qq
F_{ 32}(z)
={i\ol{\wh V_n}\/4(\pi n)^3}+{O(1)\/n^5}
\]
as $z=(\l_n^\pm)^{1\/4},n\to+\iy$.
\end{lemma}

\no {\bf Proof.}
i) Asymptotics \er{4g.aswtG} and identity \er{4g.wtF}
give \er{4g.aswcF} and the asymptotics
\[
\lb{4g.asG1i}
F(z)=\1_{4}+\cG(1,z)-\cG(0,z)+O(z^{-6})\qq\as\qq
|z|\to\iy,\qq z\in S,
\]
where $\cG=G_1=K\1_4$.
Identity \er{4g.dcLi} implies
\[
\lb{4g.wtG1a}
\begin{aligned}
\cG_{\ell j}(1,z)=
\ca \int_0^1 e^{iz\int_s^1(\xi_\ell(t,z)-\xi_j(t,z))dt}Q_{\ell j}(s,z)ds,&\ell <j\\
\qqq\qqq\qqq 0 \ \ ,&\ell \ge j\ac,
\\
\cG_{\ell j}(0,z)=
\ca\qqq\qqq\qqq 0\ \ ,&\ell <j\\
-\int_0^1 e^{-iz\int_0^s(\xi_\ell(t,z) -\xi_j(t,z))dt}Q_{\ell j}(s,z)ds,
&\ell \ge j\ac.
\end{aligned}
\]

Let $j\in\N_4$. Substituting \er{4g.wtQ} into \er{4g.wtG1a} and using \er{4g.wtQ0},
\er{Q1}, the identities
$\int_0^1q(t)dt=\int_0^1p''(t)dt=\int_0^1p(t)p'(t)dt=0$
and \er{4g.O-AB}, we obtain
$$
\cG_{jj}(0,z)=-\int_0^1Q_{jj}(s,z)ds=-{4i\/z^3}\int_0^1p^2(t)(B_2)_{jj}dt+{O(1)\/z^5}
=-{i\o_j\|p\|^2\/8z^3}+{O(1)\/z^5}
$$
as $|z|\to\iy,z\in S$.
Substituting this asymptotics and the identity
$\cG_{jj}(1,z)=0$ into \er{4g.asG1i} we obtain \er{4g.wtf2i}.

Let $z=(\l_n^\pm)^{1\/4}$ and let $n\to+\iy$. Then asymptotics
\er{4g.T2-2} gives
$
z=\pi n-{p_0+o(1)\/2\pi n}.
$
Substituting   \er{4g.wtO}, \er{4g.wtQ} into \er{4g.wtG1a}
and integrating by parts we obtain
\[
\lb{4g.aswtG1+1}
\cG_{23}(1,z)=
\int_0^1 e^{2iz(1-s)}
e^{{i\/z}\int_s^1p(t)dt}Q_{\ell j}(s,z)ds
={\cG_1^+\/z^3}+{\cG_2^+\/z^4}+{O(1)\/n^5},
\]
where
\[
\lb{XY+}
\cG_1^+=\int_0^1 e^{2iz(1-s)}(Q_0)_{23}(s)ds,
\qqq
\cG_2^+
=i\int_0^1 e^{2iz(1-s)}(Q_0)_{23}(s)\int_s^1p(t)dtds.
\]
Similarly,
\[
\lb{4g.aswtG1-1}
\cG_{32}(0,z)=
-\int_0^1 e^{2izs}
e^{{i\/z}\int_0^sp(u)du}Q_{32}(s,z)ds
={\cG_1^-\/z^3}+{\cG_2^-\/z^4}+{O(1)\/n^5},
\]
where
\[
\lb{XY-}
\cG_1^-=-\int_0^1 e^{2isz}(Q_0)_{32}(s)ds,
\qqq
\cG_2^-=i\int_0^1 e^{2isz}(Q_0)_{32}(s)\int_0^sp(t)dtds.
\]
Substituting identities  \er{4g.idA}, \er{4g.idB}, \er{4g.O-AB}
into \er{4g.wtQ0} we obtain
$$
(Q_0)_{23}=-(Q_0)_{32}=-{iV\/4}.
$$
Substituting these identities into \er{XY+}, \er{XY-}
we have
\[
\lb{X+}
\begin{aligned}
\cG_1^+=-{i\/4}\int_0^1 e^{2iz(1-s)}V(s)ds
=-{i\/4}\int_0^1 e^{-2i\pi ns-(1-s){p_0+o(1)\/2\pi n}}V(s)ds
\\
=-{i\/4}\int_0^1 e^{-2i\pi ns}
\Big(1-(1-s){p_0+o(1)\/2\pi n}\Big)V(s)ds+{O(1)\/n^2},
\end{aligned}
\]
\[
\lb{X-}
\cG_1^-={i\/4}\int_0^1 e^{2izs}V(s)ds
={i\/4}\int_0^1 e^{2i\pi ns}
\Big(1-s{p_0+o(1)\/2\pi n}\Big)V(s)ds+{O(1)\/n^2},
\]
which yields
$$
\cG_1^+=-{i\wh V_n\/4}+{o(1)\/n},
\qqq
\cG_1^-={i\ol{\wh V_n}\/4}+{o(1)\/n}.
$$
Similarly, $\cG_2^\pm=o(1)$, which yields
$$
\cG_{23}\big(1,z\big)
=-{i\wh V_n\/4(\pi n)^3}+{o(1)\/n^4},\qqq
\cG_{32}\big(0,z\big)
={i\ol{\wh V_n}\/4(\pi n)^3}+{o(1)\/n^4}.
$$
Substituting
these asymptotics into \er{4g.asG1i} we obtain \er{4g.asF3223}.

ii) Let $q,q',p,p'''\in L^1(\T)$ and let
$z=(\l_n^\pm)^{1\/4},n\to+\iy$. The integration
by parts in \er{X+}, \er{X-} gives
$$
\cG_1^+=-{i\wh V_n\/4}+{O(1)\/n^2},
\qqq
\cG_1^-={i\ol{\wh V_n}\/4}+{O(1)\/n^2}.
$$
Similarly, $G_2^\pm=O(n^{-1})$, which yields
$$
\cG_{23}\big(1,z\big)
=-{i\wh V_n\/4(\pi n)^3}+{O(1)\/n^5},\qqq
\cG_{32}\big(0,z\big)
={i\ol{\wh V_n}\/4(\pi n)^3}+{O(1)\/n^5}.
$$
Substituting
these asymptotics into \er{4g.asG1i} we obtain  \er{4g.asF3223i}.
$\BBox$

\section{Eigenvalue asymptotics}
\setcounter{equation}{0}

In this Section we will prove our main results.
Introduce the function
$$
D(\t,\l)=\det(M(1,\l)-\t \1_{4}),\qqq (\t,\l)\in\C^2.
$$
The eigenvalues $\l_n^\pm$ satisfy the identity
$D((-1)^n,\l_n^\pm)=0$ for all $n\in\N$ large enough, see \cite{BK1}.

\no {\bf Proof of Theorem \ref{4g.thaspi}.}
i) Identity \er{4g.reprM} gives
\[
\lb{asdM1}
\begin{aligned}
D(\t,z^4)=\det\big(F(z) e^{i zv(z)}-\t\1_{4}\big)
\\
=\det\ma
F_{11}(z) e^{iv_1(z)z}-\t&F_{12}(z)e^{iv_2(z)z}
&F_{13}(z)e^{iv_3(z)z}&F_{14}(z)e^{iv_4(z)z}\\
F_{21}(z) e^{iv_1(z)z}   &F_{22}(z)e^{iv_2(z)z}-\t
&F_{23}(z)e^{iv_3(z)z}&F_{24}(z)e^{iv_4(z)z}\\
F_{31}(z) e^{iv_1(z)z}   &F_{32}(z)e^{iv_2(z)z}
&F_{33}(z)e^{iv_3(z)z}-\t &F_{34}(z)e^{iv_4(z)z}\\
F_{41}(z) e^{iv_1(z)z}   &F_{42}(z)e^{iv_2(z)z}
&F_{43}(z)e^{iv_3(z)z}&F_{44}(z)e^{iv_4(z)z}-\t
\am
\end{aligned}
\]
for all $(\t,z)\in\C\ts S_r$ for some $r>0$ large enough,
where $v(z)$ is given by \er{defv}.

Let $\t=(-1)^n,\l=\l_n^\pm$ and let $n\to+\iy$. Then asymptotics
\er{4g.T2-2} gives
\[
\lb{asz}
z=\l^{1\/4}=\pi n-{p_0+o(1)\/2\pi n}.
\]
Identities \er{4g.Om} imply $ v_1(z)=- v_4(z), v_2(z)=- v_3(z)$.
Moreover, \er{asz} gives
\[
\lb{4g.asw2}
e^{-i v_1(z) z}=e^{i v_4(z) z}=e^{\pi n}(1+o(1)),\qqq
 v_2(z)z=z+{ p_0\/2z}=\pi n+\d_n,\qq \d_n=o(n^{-1}),
\]
and then
\[
\lb{idv2}
e^{i v_2(z)z}=(-1)^ne^{i\d_n}.
\]
Substituting asymptotics \er{4g.aswcF}, \er{4g.asw2}
into identity \er{asdM1} we obtain
$$
D(\t,z^4)
=e^{i v_4z}\det\ma
O(e^{-\pi n})-\t&O(n^{-3})&O(n^{-3})&O(n^{-3})\\
O(e^{-\pi n})&F_{22}(z)-\t e^{-i v_2z}&F_{23}(z)&O(n^{-3})\\
O(e^{-\pi n})&F_{32}(z)&F_{33}(z)-\t e^{i v_2z}&O(n^{-3})\\
O(e^{-\pi n})&O(n^{-3})&O(n^{-3})&1+O(n^{-3})
\am.
$$
Substituting   \er{4g.wtf2i}, \er{idv2} into the last asymptotics we obtain
\[
\lb{idDd}
D(\t,z^4)=-\t e^{i v_4z}d_n,
\]
where
\[
\lb{asdn}
d_n=\det\ma
1-e^{-i\d_n}+{i\|p\|^2\/8z^3}+O(n^{-5})
&F_{23}(z)&O(n^{-3})\\
F_{32}(z)&1-e^{i\d_n}-{i\|p\|^2\/8z^3}+O(n^{-5})&O(n^{-3})\\
O(n^{-3})&O(n^{-3})&1+O(n^{-3})
\am+O(e^{-\pi n}).
\]
Asymptotics $\d_n=o(n^{-1})$ implies
\[
\lb{iddn}
d_n=\Big(2\sin{\d_n\/2}+{\|p\|^2\/8z^3}\cos{\d_n\/2}\Big)^2
+{O(1)\/n^6}.
\]

Identity $D((-1)^n,z^4)=0$ gives $d_n=0$.
Asymptotics \er{iddn} implies $\d_n=O(n^{-3})$.

Then we have the asymptotics
$$
e^{i\d_n}+{i\|p\|^2\/8z^3}=e^{i\wt\d_n}+{O(1)\/n^6},\qqq\where\qq
\wt\d_n=\d_n+{\|p\|^2\/8z^3}.
$$
Asymptotics \er{asdn} yields
\[
\lb{4g.asF3i}
d_n=\Big(4\sin^2{\wt\d_n\/2}
+{O(\wt\d_n)\/n^5}+{O(1)\/n^{10}}\Big)
\Big(1+{O(1)\/n^3}\Big)
-F_{23}(z)F_{32}(z)+{O(\wt\d_n)\/n^6}+{o(1)\/n^9},
\]
where we used \er{4g.asF3223} and $\wh V_n=o(1)$.

Substituting asymptotics \er{4g.asF3223} into
\er{4g.asF3i} and using $\wt\d_n=O(n^{-3})$ we obtain
$$
d_n=4\sin^2{\wt\d_n\/2}-{|\wh V_n|^2\/16(\pi n)^6}+{o(1)\/n^7}.
$$
Identity
$
d_n=0
$
gives
\[
\lb{est2}
4\sin^2{\wt\d_n\/2}={|\wh V_n|^2\/16(\pi n)^6}+{o(1)\/n^7}.
\]

Let $\ve\in\C$ and let $[-i\ve,i\ve]$ be the segment of the line.
Define the single valued analytic function
$f(w)=(w^2+\ve^2)^{1\/2}-w$ in the domain
$\C\sm[-i\ve,i\ve]$ by the condition
$\lim_{|w|\to\iy}f(w)=0$.
The maximal value of the function
$|f(w)|$ on the segment $[-i\ve,i\ve]$ is equal to $|\ve|$.
Using the Maximum Principle
we obtain the estimate
\[
\lb{4g.eka}
|f(w)|=|(w^2+\ve^2)^{1\/2}-w|\le|\ve|\qqq\forall\qq w\in\C.
\]

Substituting $w={|\wh V_n|\/4(\pi n)^3}$ and $\ve^2=o(n^{-7})$
into inequality \er{4g.eka} we obtain
$$
\lt(\Big({|\wh V_n|\/4(\pi n)^3}\Big)^2+o(n^{-7})\rt)^{1\/2}
-{|\wh V_n|\/4(\pi n)^3}={o(1)\/n^{7\/2}}.
$$
Asymptotics \er{est2} yields
$$
2\sin{\wt\d_n\/2}
=\pm{|\wh V_n|\/4(\pi n)^3}+{o(1)\/n^{7\/2}}.
$$
This asymptotics implies
$$
\d_n=-{\|p\|^2\/8z^3}\pm{|\wh V_n|\/4(\pi n)^3}+{o(1)\/n^{7\/2}},
$$
then \er{4g.asw2} yields
$$
 v_2z=\pi n-{\|p\|^2\/8z^3}\pm{|\wh V_n|\/4(\pi n)^3}+{o(1)\/n^{7\/2}}.
$$
Identity \er{4g.asw2} gives
$$
z+{ p_0\/2z}=\pi n-{\|p\|^2\/8z^3}\pm{|\wh V_n|\/4(\pi n)^3}+{o(1)\/n^{7\/2}}.
$$
Then
$$
z=\pi n-{ p_0\/2\pi n}-{\|p\|^2+2 p_0^2\/8(\pi n)^3}
\pm{|\wh V_n|\/4(\pi n)^3}+{o(1)\/n^{7\/2}},
$$
which yields \er{4g.T2-2i}.

ii) Let $p,p''',q,q'\in L^1(\T)$ and let $n\to+\iy$.
Substituting \er{4g.asF3223i} into \er{4g.asF3i}
and using the asymptotics $\wh V_n=o(n^{-1})$
we obtain
$$
d_n=4\sin^2{\wt\d_n\/2}
-{|\wh V_n|^2\/16(\pi n)^6}+{o(1)\/n^9}.
$$
Repeating the previous arguments we obtain
$$
\d_n=-{\|p\|^2\/8(\pi n)^3}\pm{|\wh V_n|\/4(\pi n)^3}
+{o(1)\/n^{9\/2}},
$$
which yields \er{4g.T2-2ii}.
$\BBox$

{\bf The perfect square case.} Consider the operator $h^2$, where the operator
$h=-\pa ^2-p$ acts on the functions, satisfying
the 2-periodic conditions
$f(0)=f(2),f'(0)=f'(2)$.
The operator $h^2-\|p\|^2$ is equal to the operator
$H$ with $q=p''+p^2-\|p\|^2$. Let $p,p''\in L^1(0,1)$
and let
$\a_0^+<\a_1^-\le\a_1^+<...$ be
eigenvalues of the operator $h$.
It is well known (see \cite[Th~1.5.2]{M}), that
$$
\a_n^\pm=(\pi n)^2-p_0+{\|p\|^2-p_0^2\/(2\pi n)^2} \pm{|\wh
p_n''|+o(n^{-{1\/2}})\/(2\pi n)^2} \qqq\as\qq n\to+\iy.
$$
Therefore, the eigenvalues $\l_n^\pm=(\a_n^\pm)^2-\|p\|^2$ of the
operator $h^2-\|p\|^2$ satisfy
$$
\l_n^\pm=(\pi n)^4-2p_0(\pi n)^2-{\|p\|^2-p_0^2\/2}
\pm{|\wh p_n''|\/2}+{o(1)\/n^{1\/2}} \qqq\as\qq n\to+\iy.
$$
This asymptotics shows that asymptotics \er{asE}
is not correct.

 \setlength{\itemsep}{-\parskip} \footnotesize
 \no
{\bf Acknowledgments.} {Various parts of this paper were written
during Evgeny Korotyaev's stay in Aarhus
University, Denmark. He is grateful to the institute for the
hospitality. His study was supported by the Ministry of education
and science of Russian Federation, project 07.09.2012  No 8501 and
the RFFI grant "Spectral and asymptotic methods for studying of the
differential operators" No 11-01-00458 and partly supported by the
Danish National Research Foundation grant DNRF95 (Centre for Quantum
Geometry of Moduli Spaces - QGM)”.
Andrey Badanin's study was supported by the Ministry of education
and science of Russian Federation, project No 1.5711.2011.}

\end{document}